\documentclass[%
  superscriptaddress,
  twocolumn,
  10pt,
  aps,
  pre,
  nofootinbib, 
  final,letterpaper
]{revtex4-2}

\usepackage{graphicx}

\usepackage[utf8]{inputenc}
\usepackage{float}
\usepackage{amsmath,amsfonts,amssymb}
\usepackage[usenames,dvipsnames]{xcolor}
\usepackage{comment}
\usepackage{booktabs}
\usepackage[caption=false]{subfig}
\usepackage{microtype}
\usepackage{hyperref}

\begin{document}

\title{Homophily Within and Across Groups}

\author{Abbas K. Rizi}
\affiliation{DTU Compute, Technical University of Denmark, Kongens Lyngby, Denmark}
\affiliation{Center for Social Data Science, University of Copenhagen, Denmark}
\affiliation{Department of Computer Science, School of Science, Aalto University, Espoo, Finland}
\author{Riccardo Michielan}
\affiliation{Faculty of Electrical Engineering, Mathematics and Computer Science, University of Twente, Enschede, Netherlands}
\affiliation{Gran Sasso Science Institute, L'Aquila, Italy}
\author{Clara Stegehuis}
\affiliation{Faculty of Electrical Engineering, Mathematics and Computer Science, University of Twente, Enschede, Netherlands}
\author{Mikko Kivelä}
\affiliation{Department of Computer Science, School of Science, Aalto University, Espoo, Finland}

\date{\today}
\begin{abstract}
Homophily—the tendency of individuals to interact with similar others—shapes how networks form and function. Yet existing approaches typically collapse homophily to a single scale, either one parameter for the whole network or one per community, thereby detaching it from other structural features. Here, we introduce a maximum-entropy random graph model that moves beyond these limits, capturing homophily across all social scales in the network, with parameters for each group size. The framework decomposes homophily into within- and across-group contributions, recovering the stochastic block model as a special case. As an exponential-family model, it fits empirical data and enables inference of group-level variation of homophily that aggregate metrics miss. The group-dependence of homophily substantially impacts network percolation thresholds, altering predictions for epidemic spread, information diffusion, and the effectiveness of interventions. Ignoring such heterogeneity risks systematically misjudging connectivity and dynamics in complex systems.
\end{abstract}
\maketitle

\section{Introduction}
\label{sec:intro}
Homophily—the tendency for like to connect with like—shapes both the structure and function of networks~\cite{mcpherson2001birds, newman2002assortative, newman2003structure, noldus2015assortativity, cantwell2019mixing}. In social systems, people tend to connect with others who share attributes such as gender, age, race, education, occupation, sexual preferences, socioeconomic background, or vaccination status~\cite{moody2001race, beyrer2012global, staras2009sexual, kossinets2009origins, endo2022heavy, griggs2022role, worby2015relative, worby2015examining, hiraoka2022herd, metcalf2013persistence, aral2009distinguishing, volker2025s}. Assortative mixing occurs in various settings, ranging from friendships and workplace ties to online platforms, peer influence networks~\cite{kalmijn1998intermarriage, lewis2012social, bakshy2015exposure, tacchi2024keep}, and even graph neural networks~\cite{pei2020geom, zhu2020beyond, ma2021homophily, zheng2024missing}. It shapes interactions and governs processes like contagion, diffusion, and coordination~\cite{rizi2022epidemic, rizi2024effectiveness, hiraoka2023strength, hebert2020macroscopic, hebert2022source}. In healthcare settings, the close contact between healthcare workers and vulnerable patients can alter outbreak patterns, as homophily interacts with variations in infectiousness and susceptibility, reshaping epidemic dynamics~\cite{nguyen2020risk, shirreff2024assessing, temime2021conceptual, wallinga2006using, anderson1991infectious, geismar2024sorting, Rizi2024PhD, cure2025exponential}.

Homophily varies by interaction type and tie strength~\cite{granovetter1973strength, granovetter1983strength, mou2025quantifying}. Strong ties tend to form within close-knit groups, while weak ties more often bridge distant or dissimilar clusters~\cite{rapoport1957contribution, granovetter1983strength}. Networks of citations and trade partnerships typically exhibit dual mixing patterns, combining dense local collaboration with broad cross-group exchange~\cite{porter2009science, serrano2007patterns}. Gender homophily, in particular, is sensitive tothe interaction context and size. On Instagram, women show strong same-gender preferences in comments, while men are largely neutral~\cite{pignolet2024gender}. In face-to-face proximity networks, women tend to form homophilic triads, whereas men favor same-sex dyads~\cite{gallo2024higher}. In contrast, the teen-focused Spanish social network Tuenti shows the opposite: women sustain stronger same-gender dyads, and men more often form all-male triads than expected under null models~\cite{laniado2016gender}.

These observations suggest that homophily is influenced by both group composition and size, underscoring its variability across different social scales~\cite{peel2018multiscale, sajjadi2024unveiling}. Small, cohesive groups often display high local homophily, whereas long-range connections mix more randomly, producing distinct within- and across-group connectivity essential for understanding collective behavior \cite{granovetter2005impact}.  Nevertheless, many network models simplify homophily into a single, uniform parameter—essentially a summary statistic that captures only the network’s average mixing pattern~\cite{bojanowski2014measuring, rizi2022epidemic, hiraoka2022herd, battiston2021physics, apollonio2022novel, veldt2023combinatorial, karimi2023inadequacy}. On the other hand, models with explicit community structure, such as the Stochastic Block Model, allow different homophily values across communities but typically assign each node to a single group (and thus a single effective length scale), so they cannot capture overlapping, multiscale affiliations in which a node simultaneously participates in groups of different sizes. Epidemiological studies emphasize the importance of this complexity, demonstrating that high within-group homophily can elevate epidemic thresholds, while increased cross-group interactions typically lower them~\cite{moody2001race, grassly2008mathematical, goodreau2009birds, hebert2013percolation, burt2018structural, hebert2020beyond}. Recent studies further show that variations in homophily across interaction sizes critically influence how quickly different groups gain access to information \cite{Laber2025AccessInequality}. Therefore, to better understand network formation and dynamics, we should model homophily as a heterogeneous property that varies across social scales.

We introduce a network model that allows for homophily to operate across the network scale and scales of various finite social groups. These groups are operationalized by cliques, which serve as natural building blocks of human interactions. The cliques can represent various social foci~\cite{feld1981focused} and social circles \cite{simmel1955conflict,granovetter1973strength}, which can consist of families, workplaces, organizations, friendship groups, and other social, physical, legal, or psychological entities. In our framework, we model the homophily within cliques such that the level of homophily depends on the size of the cliques. This allows, for example, our model to have different homophily for interactions within groups and bridging groups (cliques with two nodes). By adopting a maximum-entropy formulation, the model captures homophilic structure beyond pairwise links and distinguishes between within- and across-group connections.
 This reveals scale-dependent patterns often overlooked by standard measures and aligns well with empirical studies. Our results show that such multi-scale homophily can raise or lower the percolation threshold, depending on how intra- and inter-group links are distributed. These findings offer new tools for understanding epidemic thresholds, designing interventions, and modeling the spread of information or behavior in social systems.

The remainder of the paper is organized as follows. In Sec.~\ref{sec:net_model}, we present the generative network model and demonstrate how it encodes multi-scale homophily. In Sec.~\ref{sec:results}, we validate the model against empirical data, illustrating how distinct within- and across-group mixing patterns emerge naturally within this unified framework. We then investigate how these multi-scale homophily patterns affect percolation thresholds and network connectivity, highlighting implications for epidemic control strategies and other contagion processes.

\begin{figure*}
\includegraphics[width=\linewidth]{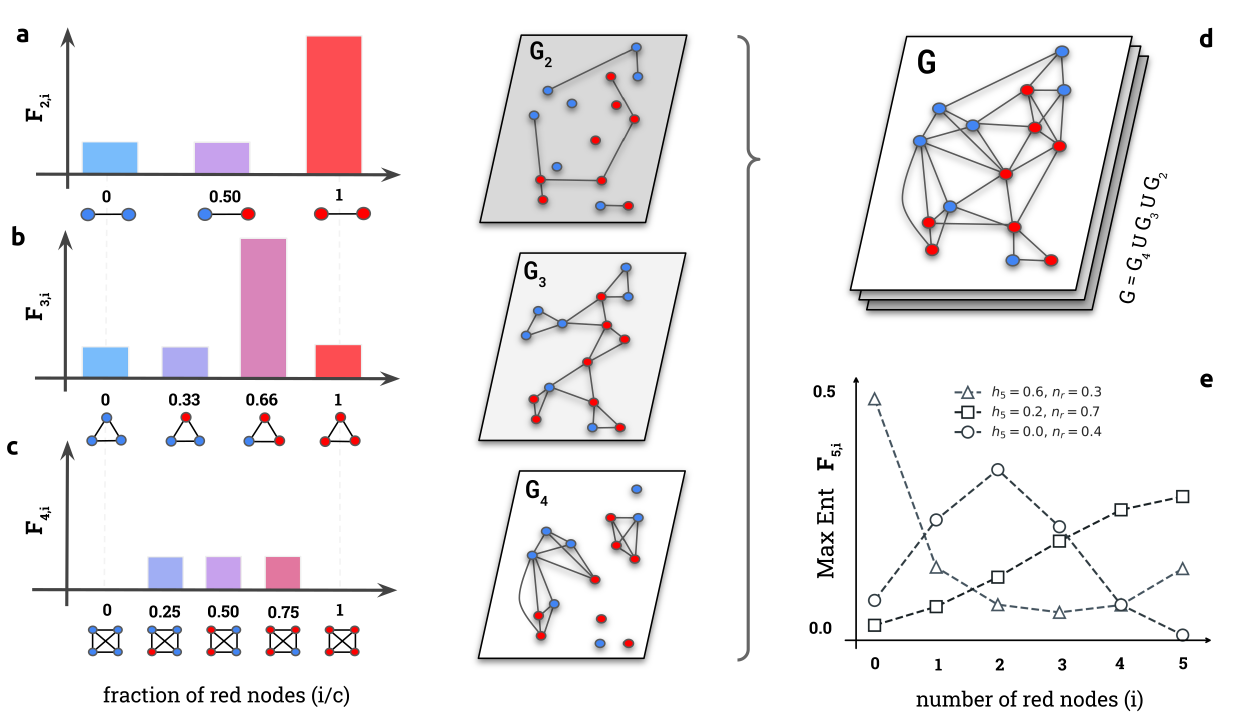}
    \caption{\textbf{Constructing a Homophilic Clique Network with Eight Red and Seven Blue Nodes.} With two colors, each $c$-clique can appear in \( c+1 \) different compositions, distributed according to \({F}_c \).
 \textbf{(a-c)} The network \( G \) is constructed by merging multiple clique layers—specifically, 2-, 3-, and 4-cliques—where each layer \( G_c \) contains cliques of size \( c \) but shares the same set of nodes. While the nodes remain fixed, each layer differs in clique size \( c \) and composition distribution \( {F}_c \). The accompanying histograms show the frequency of each \( c \)-clique type in the network \( G_c \). In each layer, \( M_c \) groups of \( c \) nodes are sampled with replacement from the available colored nodes and converted into \( c \)-cliques. For example, \( G_3 \) consists of seven sampled 3-cliques, while \( G_2 \) is constructed similarly, but with its own parameters \( M_2 \) and \( {F}_2 \). 
\textbf{(d)} All layers are then merged to form the final network: \( G = \bigcup G_c \).  
\textbf{(e)} The maximum-entropy clique composition distribution \( {F}_5 \), derived from Eq.~\eqref{eq:max_ent_f}, is illustrated for cliques of size \( c = 5 \) at homophily levels \( h_5 =0.0, h_5= 0.2 \) and \( h_5 = 0.6 \), across several values of the red-node fraction \( n_\mathrm{r} \). For clarity, we omit the corresponding networks generated from these distributions. More details in Sec.~\ref{sec:net_model}.
}
    \label{fig:schem}
\end{figure*}

\section{Model Description}
\label{sec:net_model}
We introduce a generative maximum-entropy model that both fits empirical data and provides mechanistic insight, while serving as a statistical framework for inferring group-level variations in homophily. The model is a random graph construction designed to capture homophily across multiple social scales (clique sizes). Rather than enforcing node-specific degree constraints, we fix only the average degree, allowing node degrees to fluctuate around the mean, yielding an approximately Poisson distribution.

We begin with a binary attribute assignment: each of the $N$ nodes is labeled red or blue, with $N_{\mathrm{r}}$ red nodes and $N_{\mathrm{b}} = N - N_{\mathrm{r}}$ blue nodes. Their proportions are given by $n_{\mathrm{r}} = N_{\mathrm{r}} / N$ and $n_{\mathrm{b}} = 1 - n_{\mathrm{r}}$. All subnetworks $\{G_c\}$ are defined on this same node set, so red and blue labels are consistent across layers. (The framework extends to multiple attribute classes in Sec.~\ref{sec:morecolors}.)
 We classify the cliques within each subnetwork \( G_c \) based on the number of red or blue nodes they contain. A \( c \)-clique can be classified into one of \( c+1 \) types, that are specified by the number \( i \) of red nodes included. We denote by \( F_{c,i} \) the proportion of \( c \)-cliques in \( G_c \) that contain exactly \( i \) red nodes and \( c-i \) blue nodes. To ensure that the overall fraction of red and blue nodes is preserved in the large-network limit, the distributions \( \{F_{c,i}\} \) must satisfy
\begin{equation}
   \sum_{i=0}^c \frac{i}{c}\,F_{c,i} = n_{\mathrm{r}} 
  ,
  \qquad
   \sum_{i=0}^c \frac{c - i}{c}\,F_{c,i}= n_{\mathrm{b}}.
  \label{eq:drawn}
\end{equation}

To construct each subnetwork \( G_c \), we independently sample \( M_c \) cliques of size \( c \) from the total population of nodes according to the distribution \( \{F_{c,i}\} \). 
Fig.~\ref{fig:schem}(a-c) illustrates the construction of three subnetworks with different clique-type distributions. 
In this framework, \(G_1\) is a trivial graph of isolated colored nodes, and \(G_2\) is a two-block stochastic block model \cite{holland1983stochastic}. The final network \(G\), as shown in Fig.~\ref{fig:schem}d, is obtained by merging all subnetworks $\{G_c\}$, where each subnetwork consists of cliques of size $c$ drawn over the same set of nodes;
\begin{equation}
    G= \bigcup_c G_c.
\end{equation}

Since cliques are sampled independently and we require the largest clique size to be finite, setting \( M_c = \mathcal{O}(N) \) ensures that each layer \( G_c \) remains sparse and that the fraction of overlapping links vanishes in the large-network limit. For more details, see Methods~\ref{sec:sparse}. Given a prescribed average degree \( \langle d \rangle \), controlling \( M_c \) allows us to regulate the number of links from each clique size to the overall network.  In practice, we sample fewer large cliques, reflecting empirical observations: larger cliques tend to be rarer in real-world networks due to cognitive and social constraints—a phenomenon known as “schisming”~\cite{egbert1997schisming,iacopini2024temporal}. A more detailed discussion regarding finiteness and the effect of overlap on empirical networks is provided in Methods~\ref{sec:dup}.

\subsection{Group Homophily Values}
A standard global measure for quantifying how nodes with categorical attributes (such as color) tend to connect is the \emph{Coleman index}, also known as the \emph{assortativity coefficient}~\cite{coleman1958relational, coleman1964introduction, newman2003mixing}. It is defined as
\begin{equation}
h = 
\frac{\sum_{k}(e_{kk} - n_k^2)}{1 - \sum_{k} n_k^2} 
=
\frac{e_\mathrm{rr} + e_\mathrm{bb} - n_\mathrm{r}^2 - n_\mathrm{b}^2}{1 - n_\mathrm{r}^2 - n_\mathrm{b}^2},
\label{eq:col_index}
\end{equation}
where \( e_{\mathrm{rr}} \) and \( e_{\mathrm{bb}} \) are the fractions of red--red and blue--blue links, and \( n_k \) is the fraction of nodes of color \( k \). 
By construction, \( |h| \leq 1 \): values near 1 signal strong homophily or segregation; \( h \approx 0 \) indicates random mixing; and negative values imply heterophily~\cite{salloum2022separating}.

In our framework, cliques are sampled independently, and each node participates in a Poisson-distributed number of cliques. 
Under these assumptions, the Coleman index~\eqref{eq:col_index} can be decomposed into contributions from each clique layer \( G_c \), giving
\begin{equation}
h = \sum_{c \ge 2} \alpha_c\, h_c,
\label{eq:htot}
\end{equation}
where \( h_c \) denotes the homophily of \( G_c \), and \( \alpha_c \) is the fraction of links contributed by layer \( G_c \). 
Consider a network \( G \) constructed from two layers, \( G_2 \) and \( G_6 \). Here, \( h_2 \) quantifies across-group homophily and \( h_6 \) captures within-group homophily, assuming the groups are defined by the 6-cliques. The total homophily becomes
$h = \alpha_2 h_2 + \alpha_6 h_6,$
with each layer’s influence determined by its corresponding \( \alpha_c \). For instance, \( \alpha_2 = \alpha_6 = 0.5 \) gives equal weight to both layers, whereas \( \alpha_2 > \alpha_6 \) results in across-group homophily dominating, and vice versa. In general, the weights \( \alpha_c \) can be computed as 
\begin{equation}
\alpha_c = \frac{c(c-1)}{N\langle d \rangle}M_c. 
\label{eq:Mc}
\end{equation}

Each \( h_c \) can further be decomposed into
the weighted average of homophily contributions \( h_{c,i} \) from all \( c \)-cliques with \( i \) red nodes:
\begin{equation}
h_c = \sum_{i=0}^c h_{c,i}\,F_{c,i}.
\label{eq:hc}
\end{equation}
\( h_{c,i} \) measures the deviation of such a $c$-clique's composition from random mixing and is defined as
\begin{equation}
h_{c,i} = 
\frac{
  \tfrac{\binom{i}{2} + \binom{c - i}{2}}{\binom{c}{2}} - n_\mathrm{r}^2 - n_\mathrm{b}^2
}{
  1 - n_\mathrm{r}^2 - n_\mathrm{b}^2
}.
    \label{eq:hci}
\end{equation}
This equation defines homophily at the clique level, following the Coleman approach.  
It is worth noting that Eq.~\ref{eq:col_index} can also be decomposed into contributions at the node or neighborhood level, where ``node congruity" measures each node's share of the overall homophily \cite{piraveenan2008local,piraveenan2012congruity,peel2018multiscale}. However, the formalism presented above provides sufficient measures for this study.

\subsection{Maximum Entropy Distribution}
\label{sec:maxent}
Since many different \( F_{c,i} \) distributions can produce the same overall homophily \( h_c \), we use a \emph{maximum entropy} approach to identify the least-biased distribution consistent with the desired group homophily \( h_c \) and the red and blue node fractions \( n_\mathrm{r}\) and \(n_\mathrm{b} \). Specifically, we choose \( F_{c,i} \) to maximize entropy subject to two constraints: (i) the average fraction of red nodes in cliques equals \( n_{\mathrm{r}} \) (Eq.~\eqref{eq:drawn}); and (ii) the resulting clique-layer homophily is \( h_c \) (Eq.~\eqref{eq:hc}). Since the network contains only two colors, satisfying constraint (i) automatically ensures that the blue-node fraction \( n_\mathrm{b} = 1 - n_\mathrm{r} \) is preserved, so only one constraint is needed to fix both.
The resulting distribution takes the exponential-family form:
\begin{equation}
   F_{c,i} = \frac{1}{Z} \exp{(\theta_\mathrm{r} i + \theta_{h_c} h_{c,i})},
   \label{eq:max_ent_f}
\end{equation}
where \( Z \) is the partition function normalizing over all \( c \)-clique compositions. The parameters $\theta_\mathrm{r}$ and $\theta_{h_c}$ emerge naturally as Lagrange multipliers in our maximum-entropy formulation \cite{sivia2006data}: $\theta_\mathrm{r}$ sets the fraction of red nodes, while $\theta_{h_c}$ tunes the homophily level $h_c$. These parameters are commonly referred to as \textit{conjugate couplings} in statistical mechanics. This approach naturally accommodates networks with different degrees of mixing and handles both symmetric and skewed attribute distributions. Fig.~\ref{fig:schem}e shows three examples from this versatile family of distributions.

Analogous to spin models on hyperedges~\cite{kardar2007statistical, wu1982potts, cardy1996scaling}, our framework can be viewed as a canonical ensemble, where clique homophily and color composition act as soft constraints—similar to inverse-temperature-like parameters in a spin system defined on a hypergraph. Each clique configuration corresponds to a microstate, weighted by a Boltzmann factor \( F \sim e^{-\mathcal{H}} \). 
In this formulation, the Hamiltonian \( \mathcal{H} \) depends on the full color composition of each clique. A \( c \)-clique thus represents a higher-order interaction among the spins (i.e., node attributes) of its \( c \) members.

 The goal of this model is not to replicate every microscopic detail (such as exact degree distributions), but rather to serve as a minimal generative null model that clearly isolates the role of multiscale homophily.  This two-color framework naturally generalizes to scenarios with multiple categories or even multidimensional homophily, allowing for the simultaneous consideration of various attributes. We explicitly present these extensions in Sections~\ref{sec:morecolors} and~\ref{sec:Multidimensional}.

\section{Results}
\label{sec:results}
Here, we first demonstrate that our model accurately captures empirical homophily patterns observed in real-world networks.
We then investigate how these biases in connectivity patterns influence dynamical processes unfolding on the networks.

\subsection{Model Validation}
\label{sec:data}
To assess how well our model captures homophilic structure in real-world networks, we analyze six datasets with categorical node attributes: the Slovak social networking site Pokec~\cite{takac2012data}, the Copenhagen Networks Study (CPH)~\cite{sapiezynski2019interaction}, the music-sharing platform Last.fm~\cite{asikainen2020cumulative}, a mobile phone call network~\cite{Onnela, asikainen2018cumulative}, the face-to-face proximity dataset SocioPatterns~\cite{cattuto2010dynamics, gallo2024higher}, and 100 Facebook friendship networks~\cite{traud2012social}. These datasets, summarized in Table~\ref{tab:dataset_summary}, span a broad spectrum of social interaction types—from digital connections to physical encounters.

For each empirical network \( G \), we extract all maximal cliques and group them by size to form distinct layers \( \{G_c\} \). As shown in Fig.~\ref{fig:schem}(a--d), this decomposition yields a family of subgraphs, each capturing interactions at a specific clique size. Some links appear in multiple layers, but this redundancy has a negligible effect on measured homophily. We fit the maximum-entropy model~\eqref{eq:max_ent_f} to the observed clique-type distributions by maximizing the likelihood, thereby estimating the parameters \( \{\theta\} \). For details on link duplication and its impact, see Sec.~\ref{sec:dup}.

\begin{figure*}
    \centering
\includegraphics[width=1.02\linewidth]{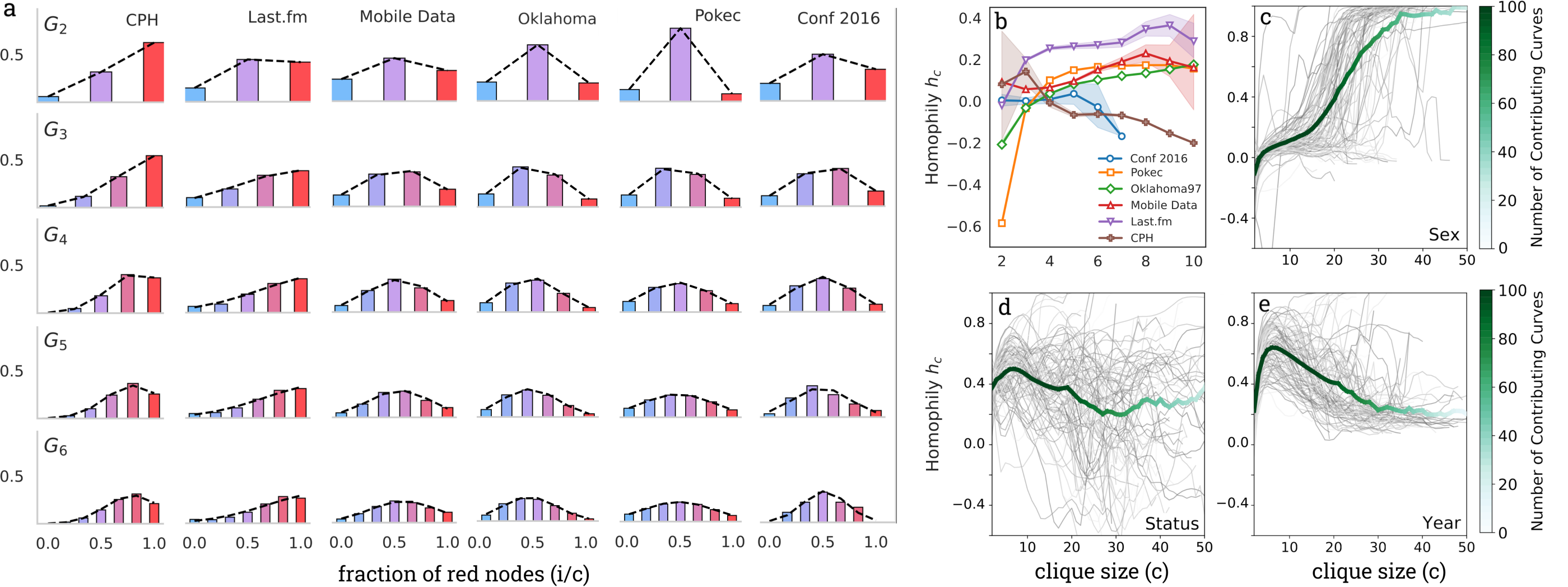}
\caption{\textbf{Empirical and Theoretical Distributions of Clique Types and Homophily in Real-world Networks mentioned in Sec.~\ref{sec:data}.} \textbf{(a)} Each column compares empirical clique-type distributions (histograms) with the corresponding maximum-entropy distributions (dashed lines, computed using Eq.~\eqref{eq:max_ent_f}), showing strong agreement and validating the model’s accuracy in capturing observed clique compositions across real-world datasets. Red represents males and blue, females. 
\textbf{(b)} Variation in homophily values \( h_c \) across clique sizes (up to \( c = 10 \)) for several empirical networks, highlighting distinct group interaction patterns based on sex attributes. Error bars represent 95\% confidence intervals obtained via bootstrap resampling.
\textbf{(c–e)} Average homophily trends across clique sizes for Facebook friendship networks from 100 U.S. institutions, with attributes grouped by: \textbf{(c)} Sex (two categories), \textbf{(d)} Student status (two categories), and \textbf{(e)} Class year (12 categories). Gray background curves represent individual institutions, while colored lines indicate the average trends. Lighter colors reflect a greater number of contributing institutions, emphasizing both inter-institutional variability and how homophily depends on the attribute under consideration.}
    \label{fig:data_real_pokec}
\end{figure*}

Our maximum-entropy approach closely matches the empirical distributions of clique compositions, as illustrated in Fig.~\ref{fig:data_real_pokec}a. Similar to Fig.~\ref{fig:schem}, it compares observed histograms of \( F_c \) across several empirical networks with the model fits from Eq.~\eqref{eq:max_ent_f}. With only two parameters—the expected homophily \( h_c \) and the fraction of nodes of one type—the model accurately captures the full distribution of clique compositions \( F_c \) in our empirical data.
This means that $h_c$ values contain all the information about the complete distributions \( F_c \) of our data, and it is enough to report the $h_c$ values to fully describe how the homophily is distributed within and across groups of different sizes.

Although these networks differ in global properties, such as degree distributions that do not necessarily align with those generated by our model, the method still effectively reproduces the patterns of homophily. This robustness stems from the fact that the model is directly constrained by empirical clique compositions or homophily values, rather than by assumptions about the degree sequence or other global network characteristics.

The Coleman indices for these datasets range from mildly homophilic (\( h = 0.1 \) in the Last.fm network and \( h = 0.2 \) in the University of Oklahoma Facebook network), to randomly mixed (\( h = 0.0 \) in the CPH and SocioPatterns networks), and mildly heterophilic (\( h = -0.2 \) on the Pokec platform). However, these aggregate measures conceal substantial variation at the clique level. We observe diverging trends across datasets. In the mildly homophilic networks, homophily increases with clique size: Last.fm exhibits random mixing for dyads (\( h = 0 \)) but stronger homophily in larger cliques (\( h = 0.3 \)); Oklahoma shows heterophily among bridging links (\( h = -0.2 \)) and homophily only in larger groups. In contrast, CPH and SocioPatterns—despite appearing randomly mixed overall—display the opposite pattern: bridging links are homophilic, but larger cliques are heterophilic.

The Pokec network offers a particularly striking example of a pattern behind an average index. Its Coleman index \( h = -0.2 \) suggests mild heterophily, yet dyads (\( G_2 \)) exhibit strong cross-sex preference with \( h_2 = -0.6 \). As clique size increases from 2 to 10, this intermingling turns into homophily. This transformation is not a minor refinement—it fundamentally changes the interpretation of the network: pairwise links reflect Pokec’s dating-oriented function, while larger cliques reflect its broader role as a social media platform where, at social scales, \textit{birds of a feather flock together }\cite{mcpherson2001birds}.

\begin{table}[b]
    \centering
    \renewcommand{\arraystretch}{1.4} 
    \begin{tabular}{l c c c r}
        \toprule
        \textbf{Dataset} & \textbf{Link Type} & \textbf{$n_\mathrm{r}$ (\%)} & $h$ & \textbf{Size} \\
        \midrule
        Mobile Phone    & Phone Calls        & 55.1  & $0.1$ & 2,173,030 \\
        Pokec           & Friendship         & 46.6  & $-0.2$ & 383,943  \\
        Last.fm         & Friendship         & 68.3  & $0.1$  & 188,672  \\
        Oklahoma FB     & Friendship         & 49.3  & $0.2$  & 16,245   \\
        CPH             & Proximity, Calls, etc. & 78.1  & $0.0$  & 779      \\
        SocioPatterns   & Proximity          & 57.4  & $0.0$  & 115      \\
        \bottomrule
    \end{tabular}
    \caption{Summary of datasets used in this study. The variable $n_\mathrm{r}$ represents the relative size of the red (male) group. The Coleman index $h$ is for overall sex homophily.}
\label{tab:dataset_summary}
\end{table}

Facebook data from 100 U.S.\ universities and colleges offers a socially comparable set of environments with rich attribute annotations. Altenburger et al.~\cite{altenburger2018monophily} previously showed that users tend to form highly sex-homogeneous friendship circles—often all-male or all-female—even though these preferences largely cancel out in aggregate, producing weak overall homophily. Fig.~\ref{fig:data_real_pokec}(c--e) illustrates how homophily varies with clique size when nodes are grouped by (c) sex, (d) student status, and (e) class year. Each attribute captures distinct social dynamics, resulting in different homophily profiles. In Fig.~\ref{fig:data_real_pokec}c, sex-based homophily increases monotonically with clique size, suggesting that sex-based clustering intensifies in larger groups. In contrast, Fig.~\ref{fig:data_real_pokec}d shows an average trend with a medium level of homophily at the link level and only a slight dependency on group size.

Our framework can analyze more complex scenarios involving multiple non-binary attributes. Fig.~\ref{fig:data_real_pokec}e reveals a clearly non-monotonic pattern for class year in the Facebook network: homophily rises in smaller cliques but levels off as groups grow, likely reflecting increased mixing between cohorts in broader social circles.  Notably, similar multiscale mixing patterns have been reported by Peel et al.\cite{peel2018multiscale}, who demonstrated that even when the network exhibits modest overall homophily, certain subgroups (such as first-year students) display markedly higher local homophily.

These observations shed a new light on
findings by Traud et al.~\cite{traud2012social}, who showed that communities in university Facebook networks tend to cluster by class year. Our analysis reveals that smaller cliques strongly reflect graduation-year homophily, but this pattern weakens in larger groups, where other attributes—such as sex or student status—become more prominent.
Similar scale-dependent patterns are well documented in the broader sociological literature. Homophily is shaped by multiple, context-dependent mechanisms~\cite{mcpherson2001birds}; academic performance influences how students reorganize their friendships~\cite{snijders2015academic}; and ethnic background plays a significant role in network formation in educational settings~\cite{boda2015ethnic}.

\subsection{Color Percolation}
\label{sec:percolation}
The structure of the network determines how robust it is to being disintegrated into small, connected components when links or nodes are removed \cite{newman2018networks, araujo2014recent, hackett2016bond, bianconi2017epidemic, sun2023dynamic}. In social systems, these removals can stem from communication failures, a limited number of contacts, or other reasons, and the connectivity then determines the spread of ideas, behaviors, or diseases~\cite {onnela2007structure, granovetter1973strength}.
Removing across-group links—often corresponding to weak ties that bridge otherwise separate communities~\cite{schawe2021network}—can trigger sudden fragmentation, severing pathways across the network. In contrast, losing within-group links, typically associated with strong local ties within groups, tends to undermine cohesion within groups without immediately disrupting global connectivity \cite{moore2000epidemics, pastor2001epidemic, newman2002spread}. This distinction is central to our analysis, where we explicitly differentiate between links that connect groups (the $G_2$-layer of our model) and those that bind them internally (the $G_c$ layers for $c>2$).

To show how group homophily values \( \{h_c\} \) influence network behavior, we analyze their effect on connectivity and percolation thresholds. Since our model generates networks with nearly identical degree distributions across different homophily settings, any changes in percolation, such as shifts in the largest connected component or the threshold, reflect structural correlations induced by homophily, not degree variation \cite{dorogovtsev2008critical}. This setup isolates the specific role of homophily within and across groups in shaping network robustness and percolation dynamics.

\begin{figure*}
\includegraphics[width=\linewidth]{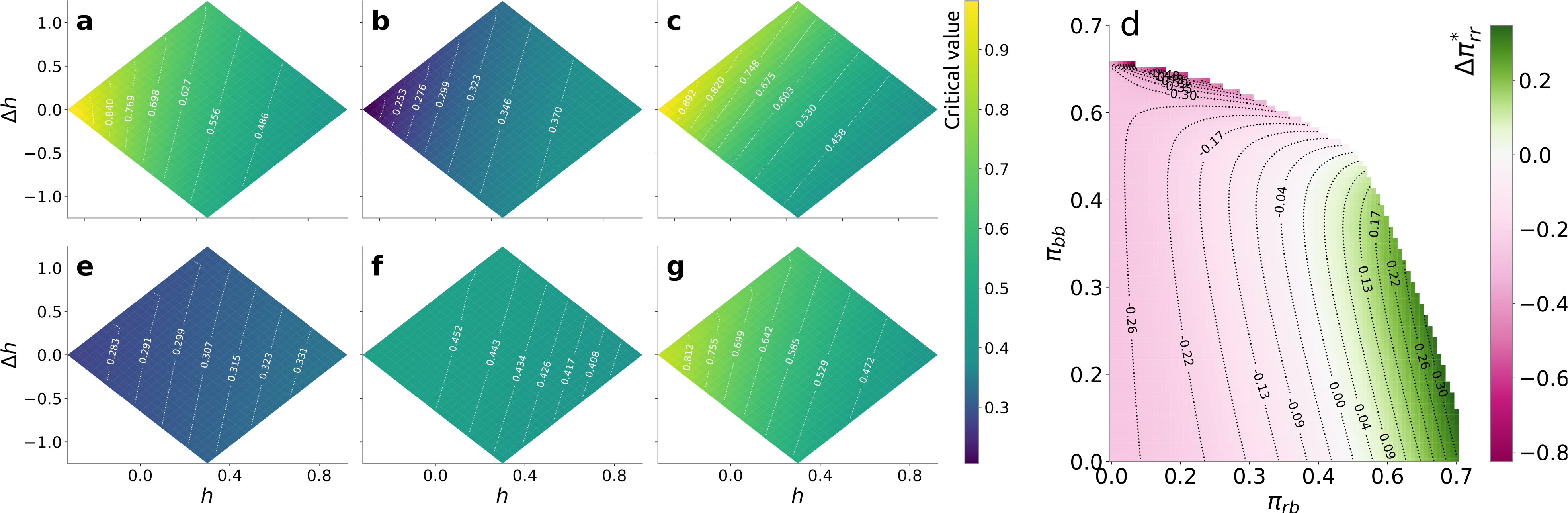}

\caption{\textbf{Within- and across-group Homophily Influences Network Connectivity, Percolation Properties and Epidemic Thresholds in Non-Uniform Ways.}
Panels~\textbf{(a–c)} present heatmaps of the critical percolation value with respect to $h=(h_2 + h_4)/2$ and $\Delta h = h_2 - h_4$  for: \textbf{(a)} \( \pi_\mathrm{rr} = \pi_\mathrm{bb} \), \( \pi_\mathrm{rb} = 0.1 \), and $\alpha_2=0.5$; \textbf{(b)} \( \pi_\mathrm{rr} = \pi_\mathrm{bb} \), \( \pi_\mathrm{rb} = 0.5 \), and $\alpha_2=0.5$; \textbf{(c)} \( \pi_\mathrm{rr} = 0.1\pi_\mathrm{bb} \) \( \pi_\mathrm{rb} = 0.1 , \alpha_2=0.2\). These panels illustrate how redistributing homophily configurations can either raise or lower the percolation threshold depending on the extent of across-group connectivity $\pi_{rb}$.
Panel~\textbf{(d)} compares the effect of redistributing homophily across small and large groups in networks with \( h = 0.5 \), average degree 2, and $\alpha_2=0.5$. Green regions indicate that emphasizing small-group homophily (\( h_4 = 0.1 \), \( h_2 = 0.9 \)) leads to a higher percolation threshold than emphasizing large-group homophily (\( h_2 = 0.1 \), \( h_4 = 0.9 \)). No percolation transition occurs in the top-right white region. Panels~\textbf{(e–g)} explore the interaction between homophily and vaccination efficacy with respect to $h=(h_2 + h_4)/2$ and $\Delta h = h_2 - h_4$. The impact of homophily depends on both within-group efficacy (\( f_v \)) and cross-group efficacy (\( f_I \)): \textbf{(e)} \( f_I = 1 \), \( f_v = 0.1 \); \textbf{(f)} \( f_I = 0.1 \), \( f_v = 0.5 \); \textbf{(g)} \( f_I = 0.1 \), \( f_v = 1 \).}
    \label{fig:app}
\end{figure*}

In a standard bond percolation process~\cite{newman2018networks}, each link in a network \( G \) is retained with probability \( \phi \) and removed with probability \( 1 - \phi \), independently and \textit{uniformly} at random. As \( \phi \) varies, we track the size of the giant component—the largest connected subgraph—to study how connectivity responds to random removal of links. This process is relevant to various simple dynamics such as epidemics, information spreading, or behavior adoption, where deleted links represent blocked transmission paths. For example, in the case of epidemic spreading, under certain assumptions, the giant component corresponds precisely to the final outbreak size~\cite{newman2002spread, hébertdufresne2025pathogendoesepidemicmake}. As the disease becomes more infectious, $\phi$ increases, and the population undergoes a continuous phase transition from a contagion‐free to an endemic state.
In more complex scenarios, link removal is not uniform and depends on node attributes. For example, disease transmission probabilities can differ based on age or vaccination status in a partially vaccinated population~\cite{goldstein2021effect, tran2021sars, roy2022modeling, mossong2008social, osei2025social, prem2017projecting, davies2020age, zhang2020changes, hiraoka2022herd}. In such cases, percolation is no longer governed by a single parameter \( \phi \), but by a vector of link-type-specific probabilities.
In our colored network model, we remove links based on a vector \( (\pi_{\mathrm{rr}}, \pi_{\mathrm{rb}}, \pi_{\mathrm{bb}}) \), where \( \pi_{kj} \) is the probability of removing a link between nodes of colors \( k \) and \( j \) (\( \pi_{kj}  = \pi_{jk}) \)~\cite{kryven2019bond}. In the large-network limit, this is equivalent to independently removing each \( kj \)-type link with probability \( \pi_{kj} \). This is similar to the idea of neighbor-induced immunity percolation~\cite{cirigliano2025neighbor}. By varying these probabilities and observing how the giant component changes, we identify when the network undergoes a phase transition in connectivity.

\subsection{Critical percolation value}
\label{sec:perc_val}
Analytical solutions for percolation on non-tree-like networks are famously intractable \cite{mezard2009information, newman2023message}; yet, by isolating within-clique spreading from across-clique behavior, our framework breaks this impasse and provides a closed-form expression for the critical percolation threshold (see Section \ref{app:critperc}). Using a multi-type branching process, we show that the percolation vector \( \boldsymbol{\pi} = (\pi_{\mathrm{rr}}, \pi_{\mathrm{rb}}, \pi_{\mathrm{bb}}) \) is critical when there exists a non-negative vector \( \boldsymbol{v} \) such that
\begin{equation}\label{eq:percmain}
    \boldsymbol{v} \in \ker \big(\boldsymbol{B}(\boldsymbol{\pi}) - \boldsymbol{I} \big), \quad \boldsymbol{v} / |\boldsymbol{v}| \geq 0,
\end{equation}
where \( \boldsymbol{I} \) is the identity matrix and the matrix \( \boldsymbol{B} = \boldsymbol{B}(\boldsymbol{\pi}) \) (see Eq.~\eqref{eq:Bpi} in Methods) encodes the expected number of nodes of a given clique type reached from another via percolation. The condition \( \det\big(\boldsymbol{B}(\boldsymbol{\pi}) - \boldsymbol{I}\big) = 0 \) is necessary but not sufficient for criticality. In expression \eqref{eq:percmain}, \( \ker(\cdot) \) denotes the kernel (null space) of a matrix—i.e., the set of vectors mapped to zero—and \( \det(\cdot) \) is the determinant, indicating when the matrix becomes singular.

We explore this condition in networks parametrized by within-group homophily \( h_4 \) and across-group homophily \( h_2 \). Fig.~\ref{fig:app}(a,b) shows heatmaps of the critical value \( \pi_{\mathrm{rr}} = \pi_{\mathrm{bb}} \) for different values of \( \pi_{\mathrm{rb}} \). When \( \pi_{\mathrm{rb}} \) is low (Fig.~\ref{fig:app}a), increasing the global homophily expressed by the Coleman index \( h \) reduces the critical threshold. When \( \pi_{\mathrm{rb}} \) is high (Fig.~\ref{fig:app}b), increasing \( h \) instead raises the threshold. Interestingly, these figures reveal how group homophilies affect connectivity, even when the Coleman index is fixed. We see that level curves for the critical value \(\pi_{\mathrm{rr}} = \pi_{\mathrm{bb}} \) are not vertically aligned, meaning that for fixed \(h\), distinct combinations of \( h_2 \) and \( h_4 \) yield different percolation thresholds. In particular, when \( \pi_{\mathrm{rb}} \) is low, percolation is more easily achieved if \( h_4 > h_2\); vice versa, when \( \pi_{\mathrm{rb}} \) is high, the percolation threshold is lower if \( h_2 > h_4\). The importance of local homophily is more prominent when $\pi_{rr}\neq\pi_{bb}$, as depicted in Fig.~\ref{fig:app}c. 
Here, the contour lines are roughly diagonal, indicating that the type of homophily is equally important to the amount of homophily for the percolation threshold. This means that the type of homophily can have a significant impact on the 
critical value of $\pi_{rr}$, which can range from 0.4 to 0.75 even when the Coleman homophily $h$ remains fixed. 

To illustrate the role of homophily within and across groups for the percolation threshold more systematically, we fix \( h = 0.5 \) with \( \alpha_2=\alpha_4 = 0.5 \) and compare two configurations: (i) weak across-group homophily (\( h_2 = 0.1 \)) and strong within-group homophily (\( h_4 = 0.9 \)), and (ii) the reverse. We compute the difference in critical values: \(\Delta \pi_{\mathrm{rr}}^{*} \).  
Fig.~\ref{fig:app}d shows that this difference can be either positive or negative, depending on \( \pi_{\mathrm{rb}} \) and \( \pi_{\mathrm{bb}} \). 
Increasing the connectivity between groups (higher \( \pi_{\mathrm{rb}} \) ) enhances the effectiveness of across-group homophily in maintaining network connectivity. In other words, when many cross-group links are present, networks with stronger across-group homophily achieve percolation at lower thresholds compared to networks with stronger within-group homophily.
Thus, stronger homophily within or across groups does not universally improve connectivity. The difference in critical \(\pi_{\mathrm{rr}}\) values between networks with the same Coleman index can be as large as 0.8, showing that global homophily alone is insufficient to predict percolation behavior.

\subsection{Epidemic Spread \& Vaccination}
\label{sec:vax}
We demonstrate how multiscale homophily influences epidemic dynamics by applying our percolation framework to contact networks. Homophily affects disease transmission through both vaccination behaviors and demographic mixing patterns. Age serves as a prime example: empirical contact matrices consistently show strong age-assortative mixing, with children primarily interacting with other children and adults exhibiting distinct interaction patterns depending on age and context \cite{mossong2008social,osei2025social,prem2017projecting,davies2020age,zhang2020changes}. Such structured contacts significantly reshape infection pathways, thereby affecting key epidemiological outcomes including herd immunity thresholds, expected epidemic sizes, and the effectiveness of interventions \cite{hiraoka2022herd,rizi2022epidemic,rizi2024effectiveness}.
Our framework facilitates a detailed analysis of intervention strategies, providing a foundation for improved design and interpretation of public health policies \cite{st2024paradoxes}. To illustrate this, we analyze a vaccination scenario involving an imperfect vaccine, categorizing individuals as vaccinated (red) and unvaccinated (blue) nodes, to investigate how homophily shapes disease spread and influences the success of interventions.

With perfect vaccines, only red–red links remain active, and any link involving at least one vaccinated node is removed from the transmission network. However, with imperfect vaccines, vaccinated individuals can still transmit the disease, though at a reduced value, to both vaccinated and unvaccinated individuals compared to the baseline probability \( \pi \) between two unvaccinated nodes.
If the across-group vaccine efficacy \( f_I \) reduces transmission between vaccinated and unvaccinated individuals, and the within-group efficacy \( f_v \) reduces transmission across vaccinated individuals, then the transmission probabilities are \( \pi_\mathrm{rb} = f_I \pi \) and \( \pi_\mathrm{bb} =  f_v \pi \), respectively \cite{hiraoka2022herd}.

Fig.~\ref{fig:app}(e–g) shows that the impact of \( f_I \) and \( f_v \) on the critical percolation threshold is non-trivial. As in Fig.~\ref{fig:app}(a–c), the effect of homophily on the critical value can reverse. In Fig.~\ref{fig:app}e, increasing either homophily raises the critical threshold, while in Fig.~\ref{fig:app}g, it lowers it. This mirrors earlier results: for low \( \pi_\mathrm{rb} \), increasing homophily decreases the critical value. Since \( \pi_\mathrm{rb} = f_I \pi \), the values of \( f_I \) in these panels explain the observed shift—\( \pi_\mathrm{rb} = \pi \) in Fig.~\ref{fig:app}e and \( 0.1\pi \) in Fig.~\ref{fig:app}g. 
Fig.~\ref{fig:app}f shows that for certain combinations of \( f_I \) and \( f_v \), the critical value becomes nearly insensitive to either form of homophily.

In tree-like networks, the required vaccine coverage for herd immunity increases with the Coleman index \(h\), and high homophily may render herd immunity unattainable~\cite{hiraoka2022herd}. Our results reveal that layered homophily creates far more complex outcomes. For example, in Fig.~\ref{fig:app}g, the critical threshold ranges from 0.5 to 0.64 when the Coleman homophily remains fixed at $h=0.5$, a 30\% difference, highlighting the influence of homophily structure on epidemic dynamics and intervention outcomes.

\section{Conclusion \& Discussion}
\label{sec:conclusion}
We introduced a generative maximum-entropy random-graph framework that captures homophily across multiple social scales with one parameter per group size. Sampling from the ensemble produces synthetic networks with prescribed multiscale homophily, and its exponential-family form enables maximum-likelihood inference of those parameters from data. Applied to diverse empirical datasets, the model integrates tightly knit groups with broader across-group interactions, accurately reproduces observed homophily patterns, and reveals structural insights that remain hidden when homophily is treated as a single global parameter. Subsequent simulation-based hypergraph work by Laber \textit{et al.}~\cite{Laber2025AccessInequality} further underscores this scale dependence, showing that varying homophily across different interaction sizes creates pronounced disparities in groups’ timely access to information—an effect explained succinctly by our analytically tractable clique-layer approach.

Additionally, we demonstrated that homophily can either increase or decrease percolation thresholds, depending on the balance between within- and across-group connectivity. This result indicates that carefully designed interventions—such as targeted vaccination or selective social distancing—can produce significantly different outcomes. Since many spreading phenomena can be effectively modeled as percolation processes, our multiscale homophily decomposition provides a principled basis for analyzing and managing diffusion processes, from accelerating information spread to combating misinformation by selectively controlling interactions within and across groups.

We note that while our model specifically addresses homophilic interactions, it omits other key structural features of social networks, such as degree heterogeneity or degree correlations. This simplification helps isolate the specific effects of group homophily, making the model particularly useful as a null model. Networks that share similar group sizes and homophily levels, yet exhibit additional characteristics like degree heterogeneity, can be compared against our model to determine whether observed differences arise from structural features beyond homophily. Thus, our framework serves to clarify whether empirical patterns are solely driven by homophily or if other structural elements also play a significant role.

The model’s exponential-family structure makes it easy to incorporate additional constraints—such as degree distributions, dynamic homophily parameters, and higher-order interactions—enabling adaptation to empirical data. It also supports extensions to multidimensional homophily, where traits interact in non-additive ways—especially when homophily in one dimension suppresses or enhances it in another~\cite{block2014multidimensional}. This versatility provides a unified framework for studying how global connectivity emerges from local assortative preferences \cite{rizi2025emergence}, offering a foundation for future work on richer dynamics, attribute interactions, and structural heterogeneity across social, biological, and technological networks.

Although our focus has been on social networks, our method can be applied to other systems with ``like-likes-like" biases. Polymer or colloidal assemblies, for example, involve monomer or particle species whose binding preferences mirror within-group homophily \cite{navarro2021microphase, karnes2020network, zeravcic2014size, navarro2021microphase}, and neural networks often exhibit connectivity shaped by neuron type or layer \cite{azulay2016c, kunin2023hierarchical}. Likewise, in gene regulatory or protein--protein interaction networks, functionally similar nodes tend to cluster \cite{yin2021emergence, rizi2021stability, navlakha2010power, apollonio2021function}. In each of these contexts, our maximum-entropy model could provide a systematic way to assess how local mixing biases influence an entire system.

Our decomposition of homophily reveals that what appears as modest or even absent assortativity at the network level often conceals a spectrum of structured, group-size-dependent mixing patterns. This observation challenges prevailing modeling assumptions and calls for a more nuanced understanding of how homophily operates in social systems, as well as the implications for the various dynamics on them. By capturing this previously hidden structure in homophily patterns, our framework not only offers better alignment with empirical observations but also provides a principled model for investigating the behavior of dynamical processes on social networks.

\section*{Acknowledgments}

\noindent \textbf{Funding:}  
AKR acknowledges support from the Independent Research Fund Denmark (EliteForsk grant to Sune Lehmann), the Carlsberg Foundation (The Hope Project), and the Villum Foundation (NationScale Social networks). RM is supported by PNRR MUR project GAMING (PE0000013, CUP D13C24000430001). CS is supported by an NWO VENI grant 202.001. MK acknowledges the grant number 349366 from the Research Council of Finland. The simulations presented above were performed using computer resources within the Aalto University School of Science's ``Science-IT'' project.

\noindent \textbf{Data Availability:}  
The datasets employed in this study were sourced from previously published research. Comprehensive descriptions of the data collection protocols, network specifications, and attribute distributions are available in the original publications. Readers are referred to the cited references for full details on each dataset.

\noindent \textbf{Code Availability:} 
Essential simulations and numerical computations are publicly available at \cite{simulations}. 

\noindent \textbf{Use of AI:}  
We utilized AI language tools to enhance grammar and readability. The authors reviewed and approved all content.

\noindent \textbf{Competing interests:}  
The authors declare no competing interests.

\bibliography{citations}

\appendix
\section*{Methods}
\subsection{Fitting the Model to Data}
Given an empirical network, we identify all maximal cliques, group them by size \( c \), and derive an empirical distribution representing the proportion of cliques containing different numbers of red nodes. Fig.~\ref{fig:data_real_pokec}a shows such distributions. Since the maximum entropy distribution for clique configurations is constrained by the observed number of red nodes per clique and the overall homophily index, we infer the latent homophily parameters \(\theta_\mathrm{r}\) and \(\theta_{h_c}\) of Eq.~\eqref{eq:max_ent_f} by maximizing the likelihood of the observed distributions.  

To ensure robust estimates, we employ a bootstrap resampling procedure for each clique size, drawing the same number of cliques 1,000 times and computing mean homophily values and 95\% confidence intervals. In Fig.~\ref{fig:data_real_pokec}b, confidence intervals follow from non-parametric bootstrap resampling of the cliques.

\subsection{Network Sparsity}\label{sec:sparse}
Let the sets of clique sizes present in $G$ be denoted by $\mathcal{C}$. The number of links in the network can be upper-bounded by
\begin{equation}\label{eq:eupperbound}
    \sum_{c\in\mathcal{C}} M_c \frac{c(c-1)}{2},
\end{equation}
as each subnetwork $G_c$ contains $M_c$ $c$-cliques. These cliques may overlap, so that Eq.~\eqref{eq:eupperbound} is an upper bound for the number of links. As sparse networks require the number of links to be $O(N)$, this sets conditions on the maximal sizes of $M_c$ and $c$. While several choices are possible to end up with a sparse network, in this manuscript we will take $M_c=O(N)$ and $\max\{c\in\mathcal{C}\}\leq K$ for some $K<\infty$. This corresponds to many small, finite groups of interactions.

\subsection{Link Duplication}
\label{sec:dup}
In large, sparse networks where cliques rarely overlap, Eq.~\ref{eq:col_index} (the Coleman index) and Eq.~\eqref{eq:hc} (the clique-based homophily formula) are equivalent. However, in finite-size empirical networks, $c$-cliques often share links, leading to redundancy. In a network \( G_c \), the number of links, denoted by \( L_c \), is always less than or equal to the total number of links in the \( M_c \) maximal \( c \)-cliques. The duplication ratio captures the extent of this redundancy
\begin{equation}
\delta_c = 1 - \frac{2L_c}{M_c c(c-1)},
\label{eq:layer_overlap}
\end{equation}
where higher \( \delta_c \) indicates greater link-sharing among cliques. 
When \(\delta_c\) is small, we can approximate the actual fraction of same-color links by expanding around the nominal (no-overlap) value:
\begin{equation}
  e_{kk}
  \;=\;
  e_{kk}^{(c)}
  \;+\;
  \gamma_{kk}\,\delta_c
  \;+\;
  \mathcal{O}(\delta_c^2),
\end{equation}
where $k \in \{\mathrm{b}, \mathrm{r}\}$ and the coefficients \(\gamma_{kk}\) capture how color-biased the overlap is: if duplication disproportionately affects red--red links, then \(\gamma_{kk}\neq0\), etc. The difference in the same color duplicated links will be
\begin{equation}
  (e_{\mathrm{rr}}
        +e_{\mathrm{bb}})
  \;-\;
  (e_{\mathrm{rr}}^{(c)}
        +e_{\mathrm{bb}}^{(c)})
  \;=\;
  \gamma \,\delta_c \;+\;\mathcal{O}(\delta_c^2),
\end{equation}
with \(\gamma = \gamma_{\mathrm{rr}}+\gamma_{\mathrm{bb}}\). 
Substituting into Eq.~\eqref{eq:col_index} yields
\begin{equation}
  h
  \;=\;
  h_c
  \;+\;
  \frac{\gamma\delta_c}{\,1 - n_\mathrm{r}^2 - n_\mathrm{b}^2\,}\,
  \;+\;\mathcal{O}\bigl(\delta_c^2\bigr).
\end{equation}
Hence, if duplication is \emph{color-unbiased} (\(\gamma=0\)), the overlap does not alter homophily at first order, implying 
\begin{equation}
\label{eq:approx_hom}
    h \;=\;  h_c \;+\; \mathcal{O}\bigl(\delta_c^2\bigr) \approx h_c.
\end{equation}
 Conversely, a nonzero \(\gamma\) indicates color-biased overlap, shifting the actual Coleman index away from the nominal \(h_c\) by roughly \(\gamma\,\delta_c\).  

When measuring the duplication ratio in an empirical network, a larger variance in node degrees can lead to a higher duplication ratio as the likelihood of the same link appearing in multiple maximal cliques increases. High-degree nodes participate in many cliques, and if their neighbors also belong to overlapping groups, shared links are counted repeatedly. In contrast, in networks where node degrees are more uniform, the chances of a link being reused across multiple cliques are lower. Additionally, if the network exhibits positive degree assortativity, where high-degree nodes preferentially connect to other high-degree nodes, this effect is further amplified, leading to even greater duplication.

\begin{table}
    \centering
    \renewcommand{\arraystretch}{1.2} 
    \setlength{\tabcolsep}{12pt}      
    \begin{tabular}{r r r r r} 
        \toprule
        $c$ & $M_c$ & $\delta_c$  & $h_c$  & $h'_c$ \\
        \midrule
        2  & 208,463 & 0.00 &  $-0.02_{-0.01}^{+0.01}$ & $0.00$ \\
        3  &  67,930 & 0.20 &  $0.20_{-0.01}^{+0.01}$  & $0.19$ \\
        4  &  25,651 & 0.31 &  $0.25_{-0.01}^{+0.01}$  & $0.22$ \\
        5  &  10,472 & 0.39 &  $0.26_{-0.01}^{+0.01}$  & $0.22$ \\
        6  &   4,293 & 0.49 &  $0.27_{-0.02}^{+0.02}$  & $0.22$ \\
        7  &   1,888 & 0.55 &  $0.28_{-0.02}^{+0.02}$  & $0.23$ \\
        8  &     824 & 0.61 &  $0.35_{-0.03}^{+0.03}$  & $0.30$ \\
        9  &     357 & 0.71 &  $0.36_{-0.05}^{+0.05}$  & $0.29$ \\
        10 &     116 & 0.66 &  $0.29_{-0.08}^{+0.09}$  & $0.33$ \\
        \bottomrule
        \label{tab:dups}
    \end{tabular}
    \caption{\textbf{Effect of Link Duplications on Homophily in the Last.fm Network.} 
    The duplication ratio $\delta_c$ is computed for the empirical data and is expected to be zero for a network generated using Last.fm statistics. The homophily values $h'_c$ and $h_c$ are computed for $G_c$ using \eqref{eq:col_index} and \eqref{eq:hc}, respectively. Error values represent bootstrap confidence intervals. The similarity between these values supports the approximation in \eqref{eq:approx_hom}.}
    \label{tab:dup}
\end{table}

Table~\ref{tab:dup} shows that real networks may have a high duplication ratio. Despite that, in practice Eq.~\eqref{eq:approx_hom} remains accurate and $h_c$ and $h'_c$ remain reasonably close. 
Eq.~\eqref{eq:col_index} is best suited for measuring homophily directly from the adjacency matrix, making it useful for assessing network-wide segregation, while Eq.~\eqref{eq:hc} provides a more refined approach for parameter inference in clique-based models, such as the maximum-entropy distribution of clique compositions.

\subsection{Aggregating Node Statuses}
\label{sec:beyond}
\noindent
Consider a discrete spectrum of colors where we aim to combine \(s\) colors into \(S<s\) colors. We can achieve this by merging certain original colors using the mapping function 
\[
\varphi : \{1, \dots, s\} \to \{1, \dots, S\},
\]
which assigns each original color \(k\) a new “merged” color \(\varphi(k)\). Say, all shades of blue to blue. After this identification, the \emph{new} distribution 
\(\widetilde{F}_{(j_1,\dots,j_S)}\) must sum all the old probabilities 
\(F_{(i_1,\dots,i_s)}\) whose indices \((i_1,\dots,i_s)\) agree with the merged color counts \(\mathbf{j}=(j_1,\dots,j_S)\).
\[
\widetilde{F}_{(j_1,\dots,j_S)}
\;=\;
\sum_{\substack{i_1+\dots+i_s = c\\[2pt]
j_\alpha \;=\; \sum_{k : \varphi(k)=\alpha} i_k}}
F_{(i_1,\dots,i_s)},
\]
This procedure \emph{lumps} together all old configurations whose new color composition is \(\mathbf{j}\). Although \(\widetilde{F}\) is a valid probability distribution, it typically does \emph{not} remain in the same exponential-family form unless every group of merged colors was already indistinguishable in the exponent. In other words, the original parameters (like \(\theta_k\)) and any homophily function must treat those merged colors \emph{identically} for the sum to factor neatly. In that \emph{renormalizable} scenario, the new \(\widetilde{F}\) preserves the same functional shape.  This coarse-graining remains an exponential family without further re-fitting if the original model assigned \emph{identical} roles to those merged categories. Otherwise, one must solve a new maximum-entropy problem in the reduced color space, and any notion of homophily now reflects only these broader, less granular labels. This is analogous to lumping the $s$-state Potts model into $S=2$ (red vs. blue), which does not necessarily result in a simple Ising model with the standard coupling \cite{wu1982potts, cardy1996scaling}.

\subsection{Beyond Two Colors}
\label{sec:morecolors}
In the definition of $h_c$ in Eq~\eqref{eq:hc} and in Section~\ref{sec:maxent}, we have focused on networks where nodes possess binary attributes, such as male and female or red and blue, for the sake of simplicity. However, this framework can be extended to nodes with discrete states or categorical attributes. Each clique network \(G_c\), and consequently the mixed clique network \(G = \bigcup G_c\), can be extended to a multi-color version.  Different colors may represent different social groups, features, or functionalities within the network. 
Assume that each node has one of \(s \geq 1\) different colors, then $\binom{c + s - 1}{s - 1}$ different $c$-clique types and hence \(s(s+1)/2\) different link types can be present.

The number of nodes of each color is denoted by $N_k$ where \(k =1,\cdots, s\), while their fraction over the total number of nodes \(N\) is denoted by $n_k$. For fixed clique size \(c\), each clique type can be uniquely specified by a vector \(\mathbf{i} = (i_1,\cdots,i_s)\) encoding the number of nodes of each color in the clique. For example, if \(s = 3\) (red, blue, or green nodes) and \(c = 4\), the vector \((2,1,1)\) corresponds to the 4-clique with 2 red nodes, 1 blue node, and 1 green node. Similarly, as in the 2-color case, different cliques yield different homophily indices,
\begin{equation}
    h_{\mathbf{i}} = \frac{\frac{ \binom{i_1}{2} + \cdots + \binom{i_s}{2}}{\binom{c}{2}} - (n_1^2 + \cdots + n_s^2)}{1 - (n_1^2 + \cdots + n_s^2)}.
\end{equation}
$M_c$ cliques are sampled according to a given distribution vector \(\mathrm{F}_c\), where \(F_{c,i}\) specifies the probability of selecting a particular clique type. Again, we constrain $G_c$ to achieve a desired network homophily value \(h\), and assume that all different color nodes have the same expected degree, similarly to Eq.~\eqref{eq:drawn}. This yields the maximum entropy clique distribution
\begin{equation}
    F_{\mathbf{i}}  = \frac{1}{Z} \exp\big(\sum_{k=1}^{s-1}\theta_k i_k+ \lambda h_{\mathbf{i}}\big).
\end{equation}

The exponents \(\theta_1,\cdots,\theta_{s-1}\) are conjugated to the constraints that ensure the average proportion of colored nodes in the cliques matches the total proportion of colored cliques. Instead, \(\lambda\) is specified after requiring the desired homophily network value \(h\). The constraints read explicitly as
\begin{align}
    N_k &= \frac{1}{c} \sum_{\mathbf{i}} i_k F_{\mathbf{i}}, \qquad k = 1,\cdots,s-1\\
    h &= \sum_{\mathbf{i}} h_{\mathbf{i}} F_{\mathbf{i}}.
\end{align}
Similarly, as in the 2-colored version, a multi-layer structured network can be obtained when mixing cliques of different sizes. 

\subsection{Multidimensional Homophily: Beyond Colors}
\label{sec:Multidimensional}
Consider nodes characterized by multiple discrete attributes (e.g., color and temperature). To describe cliques, we define a composition vector $\mathbf{I}$, where each element $\mathbf{I}_j$ indicates the number of nodes in the clique with attribute combination $j$, satisfying $\sum_{j=1} \mathbf{I}_j = c$. Under the maximum-entropy framework, the probability distribution generalizes naturally to:
\[
F_{c,\mathbf{I}} = \frac{1}{Z}\exp\left(\sum_{k}\theta_{k}\,\phi_{k}(\mathbf{I})\right),
\]
where the statistics $\{\phi_k\}$ encode relevant constraints (such as homophily per attribute or interactions between attributes), and the conjugate parameters $\{\theta_k\}$ enforce these constraints. 

\subsection{Derivation of the Percolation Threshold}\label{app:critperc}
To analyze the critical percolation value analytically, we map the percolation process on the clique network to a percolation process on a tree-like network by analyzing the spreading within the clique separately from the behavior across cliques.  We therefore calculate the average number of infected $I'$-cliques caused by a percolated node in an $I$-clique.

The average number of type $I$-cliques that are adjacent to a randomly chosen red or blue node equals
\begin{equation}
    F^*_{\mathrm{r},I}:=M_{c(I)}\frac{n_{\mathrm{r},I}F_I}{N_\mathrm{r}},\qquad F^*_{\mathrm{b},I}:=M_{c(I)}\frac{n_{\mathrm{b},I}F_I}{N_\mathrm{b}},
\end{equation}
where $n_{\mathrm{r},I},n_{\mathrm{b},I}$ denote the number of red, blue nodes in a type $I$ clique, and $M_{c(I)}$ the total number of cliques of the same size as $I$.
Furthermore, let $\boldsymbol{\pi}=[\pi_\mathrm{rr},\pi_\mathrm{rb},\pi_\mathrm{bb}]$ be the values quantifying the probabilities that red-red, red-blue, and blue-blue links are kept active after percolation. Finally, let $g_\mathrm{r}(I,\boldsymbol{\pi},n_\mathrm{r},n_\mathrm{b})$ denote the probability that, after percolation, a randomly chosen red node is still connected to $n_\mathrm{r}$ red and $n_\mathrm{b}$ blue nodes within the same clique of type $I$; similarly, $g_\mathrm{b}(I,\boldsymbol{\pi},n_\mathrm{r},n_\mathrm{b})$ denotes the probability that a randomly chosen blue node remains connected to $n_\mathrm{r}$ and $n_\mathrm{b}$ nodes within $I$.

Then, the average number of type $I'$ cliques that are reached after percolation from a node of color $(*)\in\{\mathrm{r},\mathrm{b}\}$ from a red node in a clique of type $I$ equals
\begin{equation}\label{eq:Bpi}
    B_{I_\mathrm{r},I'_{(*)}} = F_{(*),I'}^*\sum_{n_\mathrm{b}=0}^{n_{\mathrm{b},I}}\sum_{n_\mathrm{r}=1}^{n_{\mathrm{r},I}}(n_{(*)}-\delta_{\mathrm{r},(*)}) g_\mathrm{r}(I,\boldsymbol{\pi},n_\mathrm{r},n_\mathrm{b}).
\end{equation}
The term inside the inner summation is the average number of nodes of color $(*)$ in the $I$-clique that are still connected to the red node after percolation. The term in front is the average number of type $I'$ cliques adjacent to $(*)$ colored nodes. The average number of $I'$ cliques reached at a color $(*)$ node from a blue node in $I$ is denoted by $B_{I_\mathrm{b},I'_{(*)}}$, and it is calculated similarly.

Therefore, the matrix $\boldsymbol{B}= \boldsymbol{B}(\boldsymbol{\pi})$ encodes the average number of nodes of specific clique types that are reached through percolation starting within another specific clique type. Then a percolation vector $\boldsymbol{\pi}$ is critical when there is a vector $\boldsymbol{v}$, for which
\begin{equation}
    \boldsymbol{v}\in \text{ker}[\boldsymbol{B(\pi)}-\boldsymbol{I}], \quad \boldsymbol{v}/|\boldsymbol{v|}\geq 0,
\end{equation}
where $\boldsymbol{I}$ denotes the identity matrix. Note that $\text{det}(\boldsymbol{B(\pi)}-\boldsymbol{I})=0$ is a necessary condition for this to hold. \\

The elements of the matrix $\boldsymbol{B}$ require the probabilities $g_\mathrm{b}(I,\boldsymbol{\pi},n_\mathrm{r},n_\mathrm{b})$ and $g_\mathrm{r}(I,\boldsymbol{\pi},n_\mathrm{r},n_\mathrm{b})$. For any clique of type $I$ with $n_{\mathrm{r},I}$ red nodes and $n_{\mathrm{b},I}$ blue nodes, the probability that the clique remains connected after percolation is $g_\mathrm{b}(I,\boldsymbol{\pi},n_{\mathrm{r},I},n_{\mathrm{b},I}) = g_\mathrm{r}(I,\boldsymbol{\pi},n_{\mathrm{r},I},n_{\mathrm{b},I})$. This probability equals one minus the probability that there exists a cut in the clique, and can be written recursively as
\begin{align}\label{eq:pcliqueredcon}
   &g_\mathrm{b}(I,\boldsymbol{\pi},n_{\mathrm{r},I},n_{\mathrm{b},I}) = 1 - \sum_{j = 1}^{n_{\mathrm{b},I}}\sum_{i = 0}^{n_{\mathrm{r},I}}{n_{\mathrm{b},I}-1\choose j-1}{n_{\mathrm{r},I}\choose i} \nonumber \\
   &\hspace{1cm}\times g_\mathrm{b}(I_{i,j},\boldsymbol{\pi},i,j) (1-\pi_\mathrm{rr})^{i(d_\mathrm{r}-i)} \nonumber \\
   &\hspace{1cm}\times (1-\pi_\mathrm{bb})^{j(d_\mathrm{b}-j)}(1-\pi_\mathrm{rb})^{i(d_\mathrm{b}-j)+j(d_\mathrm{r}-i)},
\end{align}
\begin{align}\label{eq:pcliquebluecon}
    &g_\mathrm{r}(I,\boldsymbol{\pi},n_{\mathrm{r},I},n_{\mathrm{b},I}) = 1 - \sum_{j = 0}^{n_{\mathrm{b},I}}\sum_{i = 1}^{n_{\mathrm{r},I}}{n_{\mathrm{b},I}\choose j}{n_{\mathrm{r},I}-1\choose i-1} \nonumber\\
    &\hspace{1cm}\times g_\mathrm{b}(I_{i,j},\boldsymbol{\pi},i,j) (1-\pi_\mathrm{rr})^{i(d_\mathrm{r}-i)} \nonumber\\
    &\hspace{1cm}\times (1-\pi_\mathrm{bb})^{j(d_\mathrm{b}-j)}(1-\pi_\mathrm{rb})^{i(d_\mathrm{b}-j)+j(d_\mathrm{r}-i)}.
\end{align}

From these equations, we can compute $g_\mathrm{b}(I,\boldsymbol{\pi},n_\mathrm{r},n_\mathrm{b})$ when $n_\mathrm{r}\leq n_{\mathrm{r},I}$ or $n_\mathrm{b}\leq n_{\mathrm{b},I}$, as
\begin{align}
   &  g_\mathrm{b}(I,\boldsymbol{\pi},n_\mathrm{r},n_\mathrm{b}) = {n_{\mathrm{b},I}-1\choose n_\mathrm{b}-1}{n_{\mathrm{r},I}\choose n_\mathrm{r}}g_\mathrm{b}(I_{n_\mathrm{r},n_\mathrm{b}},\boldsymbol{\pi},n_\mathrm{r},n_\mathrm{b})\nonumber\\
   & \hspace{1cm} \times (1-\pi_\mathrm{rr})^{n_\mathrm{r}(n_{\mathrm{r},I}-n_\mathrm{r})}(1-\pi_\mathrm{bb})^{n_\mathrm{b}(n_{\mathrm{b},I}-n_\mathrm{b})}\nonumber\\
   & \hspace{1cm} \times (1-\pi_\mathrm{rb})^{n_\mathrm{r}(n_{\mathrm{b},I}-n_\mathrm{b})+n_\mathrm{b}(n_{\mathrm{r},I}-n_\mathrm{r})},
\end{align}
\begin{align}
    & g_\mathrm{r}(I,\boldsymbol{\pi},n_\mathrm{r},n_\mathrm{b}) = {n_{\mathrm{b},I}\choose n_\mathrm{b}}{n_{\mathrm{r},I}-1\choose n_\mathrm{r}-1}g_\mathrm{r}(I_{n_\mathrm{r},n_\mathrm{b}},\boldsymbol{\pi},n_\mathrm{r},n_\mathrm{b})\nonumber\\
    & \hspace{1cm} \times (1-\pi_\mathrm{rr})^{n_\mathrm{r}(n_{\mathrm{r},I}-n_\mathrm{r})}(1-\pi_\mathrm{bb})^{n_\mathrm{b}(n_{\mathrm{b},I}-n_\mathrm{b})}\nonumber\\
    & \hspace{1cm} \times(1-\pi_\mathrm{rb})^{n_\mathrm{r}(n_{\mathrm{b},I}-n_\mathrm{b})+n_\mathrm{b}(n_{\mathrm{r},I}-n_\mathrm{r})}.
\end{align}
These equations take the probability that a clique with $n_\mathrm{r}$ and $n_\mathrm{b}$ red and blue nodes remains connected from Eq.~\eqref{eq:pcliqueredcon} and~\eqref{eq:pcliquebluecon}, and multiplies it with the probability that these nodes are not connected to any remaining nodes of the clique.

\end{document}